\documentstyle[preprint,aps]{revtex}
\begin{document}
\preprint{UM-P-93/40, OZ-93/10}
\draft
\title{A Possible Way of Connecting The Grassmann Variables
And The Number of Generation}
\author{Xiao-Gang He}
\address{Research Center for High Energy Physics\\
School of Physics\\
University of Melbourne \\
Parkville, Vic. 3052 Australia}
\date{April, 1993}
\maketitle
\begin{abstract}
We construct a Left-Right symmetric
model in which the number of generation is related to Grassmann variables.
We introduce two sets of complex Grassmann variables ($\theta^1_q$,
$\theta^2_q$), ($\theta^1_l$, $\theta ^2_l$) and associate each variable
with left- and right-handed quark and lepton fields, respectively. Expanding
quark and lepton fields in powers of the Grassmann variables, we find that
there
are exactly three generations of quarks and leptons. Integrating out the
Grassmann variables, we obtain phenomenologically acceptable fermion mass
 matrices.
\end{abstract}
\newpage
How many generations of quarks and leptons are there in the Nature is one of
the
outstanding problems of particle physics today. Considerations from nuclear
synthesis\cite{1} and experimental
data on the Z decay width from LEP\cite{2} both indicate that there are
only three
generations of light neutrinos, but they do not provide information about
the number of heavy generations.
On the theoretical side the situation is not any better.
The Standard Model does not have the answer to the problem. To
answer this question one needs to go beyond the standard model. Many
theoretical efforts have been made, ranging from the topological properties
of compactified spacetimes in string theory to considerations from anomaly
cancellation and composite models, to determine
the number of generation\cite{3}. But the problem is still far
from been solved.
In this letter we will study  an interesting approach to the
generation problem which relates the particle spectrum with Grassmann
variables (GVs)\cite{4,5}. It is well known that Taylor expansions in the GVs
will have finite terms. This terminating nature of the GVs suggestes a very
interesting way to classify particle spectrum when connection between particle
fields and the GVs is made.
Extensive work has been done with models based on SU(5) grand unification
group\cite{5}.
In Ref.\cite{5} the particle spectrum, including the gauge transformation
properties, are all specified by the GVs.
In the follwoing we will study another way of relating the GVs and
the number of generation, and to construct a low energy model.

The gauge group of our model is the $SU(3)_C\times SU(2)_L\times SU(2)_R\times
U(1)_{B-L}$ Left-Right symmetric group. Under this group the
quarks $q$ and leptons $l$ tranform as
\begin{eqnarray}
q_L : (3,2,1,1/3)\;,& q_R: (3,1,2,1/3)\;,\nonumber\\
l_L : (1,2,1,-1)\;,&l_R: (1,1,2,-1)\;.
\end{eqnarray}

To make connections between the GVs and the number of generation we
introduce two sets GVs: $\theta_q = (\theta^1_q, \theta^2_q)$
and $\theta_l = (\theta^1_l, \theta^2_l)$ which transform under a global
group, $G = SU(2)_q\times SU(2)_l\times U(1)_f$ as (2,1,$\alpha$) and
(1,2,$\alpha$), respectively. We also group $q_{L,R}$ and $l_{L,R}$ into
$Q = (q_L, q^c_R)$ and $L = (l_L, l^c_R)$, and let them to transform as
$(2,1,0)$ and $(1,2,0)$, respectively. This way of
grouping the quarks and leptons suggestes that the global symmetry is in some
way related to the helicity of fermions.
The fermion and boson
fields are component fields of some super bosonic fields $E_i$ expanded in
powers of the GVs. There are two class
of expansions of the superfields. One has even powers of the GVs
and the other has odd powers. Since the overall superfields
$E_i$ are bosonic, it is clear that
the component fields with even powers of the GVs in the
expansion are
boson fileds and the ones with odd powers are fermion fields.
This expansion does not constrain the gauge transformation properties of the
component fields. The superfields can have non-trivial transformation
properties under the gauge group.

The Lagrangian density $L$ in the ordinary space-time is obtained by
first using the available superfields $E_i$ to form terms $L(E_i)$
which are singlets under both the gauge and global symmetries,
and then integrating out the GVs, that is
\begin{equation}
L =\int d^2\theta_q d^2\theta_l d^2\bar \theta_q d^2\bar\theta_l L(E_i)\;.
\end{equation}
This procedure will select certain terms in $L(E_i)$ because only the terms
with
proper $\theta$ powers will survive the integral.

Let us consider the fermion fields first. The terms with the lowest power in
the
GVs for fermion fields are
\begin{equation}
E_{1Q} = \bar \theta_q Q_1\;, E_{1L} = \bar \theta_lL_1\;.
\end{equation}
Multiplying $(\bar\theta_q\theta_q)^a(\bar \theta_l\theta_l)^b$ on $E_{1i}$, we
generate all allowed fermion fields with the same quantum numbers under the
global symmetry G in this theory. We have
\begin{eqnarray}
E_{1Q}=\bar \theta_q Q_1\;, & E_{2Q} = (\bar \theta_q \theta_q)
\bar\theta_q Q_2\;,
& E_{3Q} = (\bar\theta_l \theta_l)\bar\theta_q Q_3\;,\nonumber\\
E_{4Q} = (\bar\theta_l\theta_l)^2\bar\theta_qQ_4\;,
&E_{5Q} = (\bar\theta_q\theta_q)(\bar\theta_l\theta_l)\bar\theta_q Q_5\;,
&E_{6Q} = (\bar\theta_q\theta_q)(\bar\theta_l\theta_l)^2\bar\theta_qQ_6\;,
\nonumber\\
E_{1L} = \bar \theta_lL_1\;, &E_{2L}=(\bar \theta_l\theta_l)\bar \theta_l
L_2\;,
&E_{3L} = (\bar\theta_q\theta_q)\bar\theta_l L_3\;,\\
E_{4L} = (\bar\theta_q\theta_q)^2\bar \theta_lL_4\;,
&E_{5L} = (\bar\theta_q\theta_q)(\bar\theta_l\theta_l)\bar \theta_lL_5\;,
&E_{6L} = (\bar \theta_l\theta_l)(\bar\theta_q\theta_q)^2\bar\theta_l L_6\;.
\nonumber
\end{eqnarray}
This set of superfields carries $-\alpha$ of the $U(1)_f$ charge. One
can generate other fields with different expansions which will have different
global symmetry transformation properties.

Naively, eq.(4) contains six generations of quarks
and leptons. This is, however, not true. Some of the fields are actually the
same. To see this let us define the dual fields $\tilde E_i$. The dual of a
superfield is
defined by replacing the powers of the GVs $(\bar \theta_q)^a
(\theta_q)^b(\bar\theta_l)^c(\theta_l)^d$ to $(\bar\theta_q)^{2-b}(\theta_q
)^{2-a}(\bar \theta_l)^{2-d}(\theta_l)^{2-c}$\cite{4}.
We notice that the correct kinetic term in
the Lagrangian $L_k$ for the fermion fields can be generated by
\begin{equation}
 L_k = \int d^2\bar\theta_q d^2\bar\theta_l d^2\theta_q d^2\theta_l
\bar E_i \gamma_\mu D^\mu \tilde E_i\;,
\end{equation}
if the component field of $E_i$ is the same as the component filed of its dual.
We therefore require that the component field is the same as the component
field of its dual. It is easy to see that the following fields are dual pairs
\begin{eqnarray}
E_{1Q}\leftrightarrow E_{6Q}\;, &E_{2Q} \leftrightarrow E_{4Q}\;,
&E_{3Q}\leftrightarrow E_{5Q}\;.\nonumber\\
E_{1L}\leftrightarrow E_{6L}\;, &E_{2L}\leftrightarrow E_{4L}\;,
&E_{3L}\leftrightarrow E_{5L}\;.
\end{eqnarray}
We, therefore, have only three generations of quarks and leptons.

We now turn to possible Higgs scalars $H_i$ which may generate fermion masses
through Yukawa terms. The Yukawa terms will have the form $E_iE_jH_k$. It is
clear that in order to
couple the Higgs scalars $H_i$ to fermions,
$H_i$ should carry two times of the $U(1)_f$ charge as
$E_i$ but with opposite signs. We have
\begin{eqnarray}
H_1 =  \theta_q \theta_q h_1\;, &H_2 = \bar\theta_l \theta_l \theta_q\theta_q
h_2\;, &H_3 =(\bar \theta_l \theta_l)^2 \theta_q \theta_q h_3\;,\\
H_4 = \theta_l \theta_l h_4\;, &H_5 =\bar \theta_q \theta_q
\theta_l\theta_lh_5\;,&H_6=(\bar \theta_q \theta_q)^2\theta_l\theta_l h_6\;.
\nonumber
\end{eqnarray}
{}From our previous definition for duals, we find $H_2$ and $H_5$ are
self-dual,
and $H_3$, $H_6$ are the duals of $H_1$, $H_4$, respectively. $h_i$ are
singlets under the global symmetry G.
The gauge transformation properties are not specified. In order to form
gauge singlets with fermions to generate masses, we assign $h_i$ to transform
as $(1,2,2,0)$ under the gauge group. Notice that since $h_i$ are singlets
under the global symmetry G, we can also expaind $H_i$ by replacing $h_i$ to
 $\tilde h_i = \tau_2 h^*_i\tau_2$ without change the overall transformation
properties of $H_i$ under the gauge and the global symmetries. Therefore
$\tilde h_i$ should also been included in eq.(7).
With the fields in eq.(7), we find that only the
following Yukawa terms will survive the GV integration
\begin{eqnarray}
E_{1Q}E_{1Q}\tilde H_1\;, E_{3Q}E_{3Q}H_1\;,& E_{1Q}\tilde E_{2Q} H_1\;,
 E_{1Q}E_{3Q}H_2\;,
\nonumber\\
E_{1L}E_{1L}\tilde H_4\;,  E_{3L}E_{3L}H_4\;,& E_{1L}\tilde E_{2L}H_4\;,
 E_{1L}E_{3L}H_5\;.
\end{eqnarray}
Because both $h_i$ and $\tilde h_i$ are available to form gauge singlets with
fermions, each term in eq.(8) contains two terms. For quarks we have
\begin{equation}
E_{iQ}E_{jQ}H_k \Rightarrow \lambda_{ij}\bar q_{iL} h_k q_{jR}
+\lambda'_{ij}\bar q_{iL} \tilde h_k q_{jR}\;,
\end{equation}
where $\lambda$'s are constants.  Similarly for leptons.
When $h_i$ develop vacuum expectation
values, the quarks and leptons obtain their masses. If we now identify
\begin{eqnarray}
q_1 = (c,s)\;,&q_2 = (u,d)\;, & q_3 = (t,b)\;,\nonumber\\
l_1 = (\nu_{\mu},\mu)\;, &l_2 = (\nu_e, e)\;, &l_3 = (\nu_\tau, \tau)\;,
\end{eqnarray}
we obtain the following form for the fermion mass matrices
\begin{eqnarray}
M_f = \left(\begin{array}l
0\;\; a\;\; 0\\
a^*\;\; b\;\; c\\
0\;\;c^*\;\; d
\end{array}
\right)
\end{eqnarray}
One can have
different way to identify the first, the second and the third generations as
suggested in eq.(10). However for quarks the identification in eq.(10) is the
only one which is consistent with experimental data. For leptons there is more
flexibility. In this model neutrinos only have Dirac masses.

In the above discussions only certain superfields are studied. The complete
expansion of the superfields in powers of the GVs will generate
more particles. There will be several different classes of component
fields with different quantum numbers under the global symmetry G.
We have no contral of the gauge
transformation properties for each class of the fields. However for a given
class  of the superfields we can follow the some procedure discussed before
to determine the number of independent component fields. In this sense the
number of generation or particle spectrum is connected with the GVs.

In conclusion we have suggested a possible way to connect the GVs
and the number of generation. We constructed a model with three
generations of quarks and leptons and phenomenologically acceptable fermion
mass matrices. We must say that more studies are needed to understand
the relation betwteen the GVs and the number of
generation if they are really connected. The possibility suggested in this
paper is only one example.

\acknowledgments
I thank Prof. Delbourgo for discussions. This work was
supported in part by the Australian Research Council.

\end{document}